# Homeopathic Modernization and the *Middle Science Trap*: conceptual context of ergonomics, econometrics and logic of some national scientific case

Eldar Knar[1]

Institute of Philosophy, Political Science and Religion Studies,
Ministry of Science and Higher Education of the Republic of Kazakhstan

*https://orcid.org/0000-0002-7490-8375*
*eldarknar@gmail.com*

Abstract

This article analyses the structural and institutional barriers hindering the development of scientific systems in transition economies, such as Kazakhstan. The main focus is on the concept of the "middle science trap," which is characterized by steady growth in quantitative indicators (publications, grants) but a lack of qualitative advancement. Excessive bureaucracy, weak integration into the international scientific community, and ineffective science management are key factors limiting development. This paper proposes an approach of "homeopathic modernization," which focuses on minimal yet strategically significant changes aimed at reducing bureaucratic barriers and enhancing the effectiveness of the scientific ecosystem. A comparative analysis of international experience (China, India, and the European Union) is provided, demonstrating how targeted reforms in the scientific sector can lead to significant results. Social and cultural aspects, including the influence of mentality and institutional structure, are also examined, and practical recommendations for reforming the scientific system in Kazakhstan and Central Asia are offered. The conclusions of the article could be useful for developing national science modernization programs, particularly in countries with high levels of bureaucracy and conservatism.

Declarations:

No conflicts of interest
Article was not funded
LLM was not used

---

[1] Fellow of the Royal Asiatic Society of Great Britain and Ireland

# Introduction

Bureaucratic traditions in the system of state administration in post-Soviet countries, including those in Central Asia and Kazakhstan, are strong and deeply entrenched. Any political and civil innovation inevitably becomes entangled with bureaucratic excess, hindering progressive initiatives and the dynamics of development. The struggle against bureaucracy and corruption in such environments only yields short-term effects. In the medium and long term, the bureaucratic apparatus inevitably regenerates and adapts to new conditions. Therefore, the Hippocratic principle "*Contraria contraries curentur*" (the opposite is cured by the opposite) is of minimal effect in such systems. More often than not, it brings no result at all.

To achieve the maximum effect of real (not simulated) development, fighting bureaucracy is inefficient. However, if progressive ideas or communities can adapt to bureaucracy and play by its rules, positive societal, social, and governmental dynamics are possible. Simply put, in such conservative and bureaucratic political systems, the principle "*Similia similibus curentur*" (like cures like) can be interpreted as "Bureaucracy is cured by bureaucracy."

We focus on just one aspect of the state that also requires development and progress in a conservative environment—science. In the bureaucratic system, all areas are weak, but science is the "champion" in terms of bureaucratic weakness or, more precisely, social, public, and economic ineffectiveness (in terms of impact). At least in Central Asia and Kazakhstan.

In post-Soviet bureaucratic and conservative systems, a "sick" science requires stationary rest and should not be disturbed by excessive surgical radicalism. Therefore, for the development of national scientific systems, it is more rational and effective to use subtle tools and therapies, which would satisfy the bureaucratic council. However, the essence of therapeutic subtlety is that small changes are perceived as loyal by the conservative bureaucracy while still having a maximum effect on the development of the scientific system. We call this approach "homeopathic modernization."

Supersmall doses of transformation, change, and modernization are not strongly felt by the scientific and systemic bureaucracy, but their implementation creates a resonant effect, resulting in effective and productive development of the national scientific system.

In this article, we do not describe the current state of science in Kazakhstan (except for some necessary details). The scientific status quo is adequately covered in documents such as the annual reports of the National

Academy of Sciences of the Republic of Kazakhstan. Many assertions will rely on corresponding data, but details will be omitted by default. We also avoid vague and contentless epithets such as "need to develop," "strengthen," "increase," and so on. The subsequent content will be as rational and pragmatic as possible. If epithets are used, they will only be accompanied by precise quantitative characteristics.

Instead, we explore some interpretations of President Kassym-Jomart Tokayev's statements about creating a "high-quality and progressive science" and a national "research hub" in the Republic of Kazakhstan in the context of and through the lens of "homeopathic modernization."

We will approach this constructively. We will not claim that our science is good and just needs improvement – this is irrational and incorrect. We must assert that our science is extremely weak and that the scientific potential is close to zero, but it can be transformed into a leading "high-quality and progressive science" and a high-level "research hub." According to the plans and directives of the President of Kazakhstan, this approach is rational and correct.

Thus, for clarity and brevity, we focus on three fundamental and foundational aspects:

- Actualization of the problem
- Solution of the problems
- Prospects

Furthermore, in descriptions of Kazakh science, the problems of the scientific system are not accurately or correctly actualized.

**Literature Review**

Scientific systems, especially in transition economies such as Kazakhstan, face barriers caused by inefficiency in management, excessive bureaucracy, and a weak connection between scientific institutions and government structures. This leads to a phenomenon that can be described as the "middle science trap," where a country or organization reaches a certain level of development but cannot advance to a higher level due to institutional and administrative issues. These barriers result in stagnation of qualitative growth in scientific systems, despite quantitative success [Knar, Eldar, 2024].

One of the key factors limiting the development of science in post-Soviet countries is the high degree of state intervention in scientific activities.

This necessitates reforms in institutions, control, and the allocation of scientific resources. The role of the state in science policy was thoroughly explored by Etzkowitz and Leydesdorff [2000], who demonstrated how successful reforms in countries such as Germany and the United States allowed overcoming bureaucratic barriers and creating a more efficient model of science management integrated with industry.

Bureaucratization, primarily described by Weber [1947], shows how formal management structures can lead to stagnation in scientific systems, limiting their ability to innovate. Elman and Kaplan [2020] demonstrated that excessive control and opaque grant systems suppress scientists' initiative. Merton [1942] emphasized that bureaucracy in science creates "norms" and "regulations" that hinder the implementation of new ideas and innovative solutions. Latour [1987] and other sociologists of science [Giddens, 1984] argue that scientific institutions are shaped not only by scientists but also by administrative and governmental structures, which may limit the freedom of scientific thought.

The concept of the "middle science trap" is analogous to the "middle income trap" [Glawe, Wagner, 2016], where countries or scientific systems face issues of qualitative growth. According to the economic theory of the "middle income trap," countries that have reached a middle level of economic development cannot break into higher categories because of institutional barriers and internal structural problems. In the scientific system, this analogy holds true when countries or scientific institutions achieve stable quantitative indicators (e.g., the number of scientific publications) but cannot take science to a qualitatively new level, often owing to a lack of innovation and poor adaptation to global trends.

The connection between science and innovation remains weak in Kazakhstan. Research shows that innovations emerge as a result of high-quality fundamental research [Carroll, 2021]. However, in Kazakhstan, where the scientific system is focused on quantitative indicators, innovation activity remains low.

In Kazakhstan, according to the SCImago Journal & Country Rank [2023], the situation is characterized by stagnation: despite an increase in the number of publications, the quality of research remains low, and science does not reach international competitiveness. A similar situation is observed in other Central Asian countries, where science faces administrative barriers and political and economic problems, preventing improvements in innovation activity. These problems remain systemic and reflect the lack of institutional reforms.

North's [1990] institutional theory emphasizes that scientific systems often become stuck in inefficient structures because of dependence on past experience ("path dependency"). This is especially relevant for post-Soviet countries, where old bureaucratic management systems remain even after the transition to a market economy.

Cultural dimensions [Hofstede, 1980] also play a significant role. Bisenbayev [2023] noted that in countries with a high power distance index, such as Kazakhstan, bureaucracy is perceived as a natural element of governance. This makes it difficult to implement reforms aimed at decentralization.

The "homeopathic modernization" approach focuses on gradual changes that minimize resistance while simultaneously leading to significant improvements in the scientific system. Unlike large-scale reforms, which may provoke negative reactions from bureaucracy and stress the system, small but strategically important changes can significantly influence the efficiency of scientific institutions and improve international competitiveness [Sharma, 2020; Carroll, 2021]. Implementing minimal changes, such as digitization, improving the expertise system, and optimizing grant distribution, can be the keys to overcoming bureaucratic barriers and improving the quality of science.

International experience shows that minimal changes in scientific infrastructure can substantially improve the effectiveness of scientific work. China serves as an example of the successful implementation of such a strategy, reducing bureaucratic burdens on scientists and accelerating the adoption of new technologies [Zhang, 2018]. In India, significant improvements in research quality were achieved through the digitization of scientific activity and improvements in grant mechanisms ]Sharma, 2020]. In European Union countries, the Horizon 2020 program demonstrated how to create open and transparent mechanisms for distributing funding, which reduces administrative barriers and increases the participation of research groups [European Commission, 2021].

## Results

### *Essences*

According to the chronology of objective ranking indicators, since 2011, Kazakhstani science has been in a state of qualitative degradation or stagnation, with some growth in certain quantitative indicators. Specifically,

Kazakhstan has ranked 99th in the world Hirsch index for the tenth consecutive year (since 2014 and earlier) without any positive or negative dynamics. In terms of publication activity, Kazakhstan has not changed its position in the past ten years (66th in 2013 and 66th in 2023, with minor fluctuations).

This state of Kazakhstani science can be described as the "middle science trap," which is analogous to the well-known economic paradigm of the "middle-income trap." This analogy is natural and logical, directly resulting from identical causes and effects. If the middle-income trap is caused by specific economic relations, the middle science trap is caused by bureaucracy and bureaucratization in the context of resource scarcity and a lack of competence. At the same time, the relatively high potential of science prevents it from falling lower despite all the costs, shortcomings, and problems. However, the high level of bureaucratization and low level of management competence replace qualitative development with quantitative growth, preventing Kazakhstani science from breaking new ground.

Despite the information wave about the successes of Kazakhstani science, the real situation objectively and subjectively testifies to its critical state, with no clear signs of positive dynamics in the short, medium, or long term. Long-term staying in the realm of mediocrity (the middle science trap) inevitably leads to the degradation of science or the neutralization of its growth and development potential.

A logical question arises: Can Kazakhstani science take a prominent position in the global scientific community?

Therefore, before moving on to the actualization of problems, solutions, and prospects, we must answer the following questions:

Is the development and progress of science possible in the current political realities of Kazakhstan?

Is it possible to maximize the development of science with minimal impact on the system?

***How can we escape the "middle science trap" and create a "high-quality and progressive science" in Kazakhstan?***

Is it possible for science to develop and progress in the current political realities of Kazakhstan?

We answer with absolute certainty: Yes, it is possible!

Science is a special category and a unique type of activity. Science can develop in any system, structure, or environment. Regardless of how extractive or inclusive the system or environment is. Regardless of how liberal or conservative the system is.

In rent-oriented environments, political infrastructure develops poorly; in authoritarian environments, civil society develops poorly; and in conservative systems, social relations develop poorly. However, science is invariant, and whether it develops or stagnates depends only on the political will and state thinking of those in power.

The top 30 countries in terms of the quantitative indicator of science (total number of publications) are the USA, China, India, Russia, Iran, Brazil, Saudi Arabia, Indonesia, Malaysia, Egypt, Pakistan, Hong Kong, and others. In terms of the quality indicator (H-index), the first leadership group includes the USA, China, India, Singapore, Russia, Hong Kong, Brazil, New Zealand, and others.

These are completely different formations: liberal and authoritarian regimes, open and closed societies, progressive and conservative states, theocratic and secular states, and countries with vastly different populations—from 5 million (New Zealand) to 1.5 billion (China).

Therefore, a country can become a world leader in science regardless of the established system of economic, social, public, and political relations.

Of course, liberal systems are generally more effective in science than conservative systems are. However, this difference (as seen in the Shimago example) is not fatal, and the magnitude of these gaps is not large enough to speak of the principled impossibility of creating outstanding science in any society. We don't need to be ahead of the entire world; it's enough to be ahead of the majority.

Thus, we can draw the most important conclusion (for clarity, we divided the conclusion into three parts, but it is one overall conclusion):

The effectiveness or ineffectiveness of the grant system depends primarily on the system and structure of science management and administration (that is, on the narratives and paradigms of political will),

The effectiveness or ineffectiveness of science as a whole primarily depends on the system and structure of science management and administration,

The development or stagnation of science primarily depends on the system and structure of science management and administration.

Therefore, if the state interprets political will and state thinking regarding system narratives in a rational and pragmatic way, science in that state can become advanced or distinguished regardless of other influencing factors (institutional variables).

In simpler terms, if the state reconfigures the management system in the correct direction and appropriately, science can and should become a leading national brand in terms of global reputation, competitiveness, and "soft power."

This is the answer to the following question: Can Kazakhstani science become advanced and progressive, according to the plans and directives of the president? Yes, it can and should!

This is true if the state creates an effective, ergonomic, and adaptive system for managing and administering science.

Moreover, an effective and rational management system must address the most important task: qualitative development ahead of quantitative growth in the national scientific system.

***How can the development of science be maximized with minimal impact on the system?***

On the one hand, achieving radical improvements in science clearly requires radical methods and means. For example, switching to the Israeli model (Office of the Chief Scientist), the Soviet model (the dominance of the Academy of Sciences without a ministry, with a State Committee for Science), or the American model of direct science management through the president's office (President's Council of Advisors on Science and Technology, PCAST). Or, for instance, increasing spending on science to 3% or more, and so on.

However, such radicalism does not carry a constructive and realistic role. For example, our system and structure for administering, managing, and overseeing science is too complex, cumbersome, and inflexible. It is unlikely that we can implement measures for radical reform of this system and structure. The resistance factors are too strong. If such a form exists, it means that it is necessary. We can remove one or two links, but changing the established system and form will not work.

Thus, we cannot propose unrealistic development scenarios. Therefore, we do not consider the need for a radical reconstruction of the administration and management of science. What is needed is a minimal contribution with maximal output.

For this purpose, it is necessary to optimally and correctly prioritize powers, functions, and the decision-making process in the science sector within this system. We must calibrate, fine-tune, and adjust the existing form so that it operates as effectively and efficiently as possible.

In other words, we need a set of measures for minimal impact on the system to achieve the maximum effect in terms of results, efficiency, and development.

Here, we briefly outline the minimal measures that should be implemented in the first stage:

- Actualize, balance, and optimize the functions of the Ministry of Education and Science of the Republic of Kazakhstan (MNE RK) and the Academy of Sciences under the President of the Republic of Kazakhstan.
- Optimize the grant system.
- Optimize the legal and regulatory framework of the grant system.
- Optimized scientific processing.
- Optimize the system of scientific project expertise.

This list of measures for significant progress in science may seem unremarkable. However, in reality, this is exactly the case where "God is in the details" or "The devil is in the details" simultaneously.

The problem of science is not only, or even primarily, in major flaws but also, to no lesser extent, in small destructive details. If these details are replaced with constructive details, the level of efficiency and effectiveness of science will increase not by a few percent but several times. This assertion is based on our own experience and the analysis of other scientific projects.

***How can the "Middle Science Trap" be obtained and "High-Quality and Progressive Science" be created in Kazakhstan?***

To create a full-fledged and valuable Kazakhstani research hub, the process of development should be broken down into two key stages:

- The quality development from 2025 to 2026 is inclusive.
- Quantitative growth from 2027 onwards.

This refers to the main processes of transformation and reconstruction of the scientific system. After 2026, the process of quality development will continue in a consistent, permanent, and latent form.

From a cause-and-effect perspective, bureaucracy and bureaucratization (B&B) as causes lead to incompetence in management and corruption risks in the national scientific system as consequences.

In this context, we interpret quality development as the need to eliminate or minimize bureaucracy and bureaucratization (B&B) within the scientific system. However, the bureaucratic structure itself can remain virtually unchanged (it is necessary as a systemic and formal segment of the state and governance in the field of R&D). However, we must minimize the degree of freedom for bureaucracy and bureaucratization within it.

Minimizing the degrees of freedom for B&B will significantly increase competence and expertise in the management, administration, and management of science. It will also eliminate or minimize corruption risks and costs in the national scientific system.

We assert that this action will automatically lead to multiple increases in the effectiveness and results of Kazakhstani science—without the need for radical transformations.

In this sense, Kazakhstani science can and should become a testing ground (or pilot project) for the implementation and adoption of these principles.

Here, we must return to the role of individuals in science. The quality of the managerial personality and staff predominantly influences the efficiency and effectiveness of the scientific system. According to the definitions and provisions of the "Concept of Public Administration Development in the Republic of Kazakhstan until 2030: Building a 'Human-Centred' Model – 'People First'" [Decree of the President of the Republic of Kazakhstan, February 26, 2021, No. 522].

In science, what matters is not the structure of management but the archetype of science-oriented thinking, HIPPO (Highest Paid Person's Opinion). This is important in all sectors, especially in science. In science, directives do not work—paraphrasing a phrase from a film, "Scientific committees don't win races." In science, only conscious necessity, conceptual pragmatism, and ergonomic rationalism work. Even the Manhattan Project succeeded because of the understanding of the specifics of science and the mentality of scientists by military leadership. Unfortunately, we do not have such an understanding, or it is clearly insufficient.

In any case, the HIPPO institution in science management must consist of scientists with competencies in epistemology and gnoseology.

For quantitative growth, it makes no sense if it is not validated and verified by quality development.

For example, simple mechanical increases in the number of scientists, grants, or institutions will not lead to the "middle science trap." This is not the case where quantity transitions to quality.

All these transformations must be carried out with direct involvement and under the auspices of, primarily, the Administration of the President and the Anti-Corruption Committee. However, this participation and collaboration must be based exclusively on strict scientific methodology and a maximal understanding of scientific processing. Otherwise, we will simply replace one formality with another—hypothetical and without visible effects and results.

We have summarized practically our entire position on the development of Kazakhstani science. This is necessary for clarity and brevity in our proposals. The following content expands on some of the mentioned entities.

### *The Power of Power*

External perfection acts magically, stimulating inviolability and untouchability. Therefore, internal flaws, defects, and wear go unnoticed, and collapse occurs instantaneously.

This also applies to the scientific system in Kazakhstan. Everything looks perfect here. There are the Supreme Scientific and Technical Committee, the National Science Council, the relevant ministry, the Academy of Sciences, departments, commissions, committees, funds, competence centers, commercialization centers, excellence centers, science offices, commercialization centers, hubs, shared-use laboratories, national scientific councils, scientific councils, public councils, laws, and other regulatory legal acts. There are even scientists.

However, despite the structural perfection of this system, there is one problem: it does not work or, more precisely, it works ineffectively and suboptimally (judging by the results). It does not work primarily because there is structure and formality, but it has no real system or systemic functionality.

This has been acknowledged at the highest level. According to the president of Kazakhstan:

*"Unfortunately, science in our country has not received proper attention for many years. As a result, many unresolved issues have been identified. It can be said that this field is lagging" (kaztag.kz/ru/news/nauke-ne-udelyalos-vnimanie-i-nakopilos-mnogo-nereshennykh-voprosov-tokaev).*

Thus, the current state and potential of Kazakhstani science clearly do not meet state expectations, economic engagement, and public demand.

The primary reason for the lag in the scientific field lies in the bureaucratization of the scientific system.

The problem of causal attribution lies in some misunderstanding of what the problems of science are. Typically, when asked, "*What are the problems of Kazakhstani science*?" many provide the standard set of answers:

- low level of funding,
- too few scientists,
- weak level of expertise,
- low level of commercialization,
- weak material and technical base, and so on.

However, in reality, these are not problems. They are the consequences and the direct and indirect effects of the problem. The main and fundamental problem of Kazakhstani science is as follows:

- the high level of bureaucracy and bureaucratization of the scientific system.

Bureaucratization is the "glass ceiling" for the scientific system. It prevents science from rising upwards. Today, the level of bureaucratization in management is such that it prevents Kazakhstani science from breaking out of the "middle science trap."

Without solving the problem of excessive B&B in science, no other problem can be solved or its consequences eliminated.

Naturally, the carrier and source of bureaucratization are Bureaucracy as a class and formation. However, bureaucracy is neither good nor evil. Bureaucracy is not a moral category but a purely managerial one. It can only be effective or ineffective. The state cannot function without bureaucracy, as bureaucracy forms the basic level of systemic governance and is the primary form of the modern state.

However, when bureaucracy becomes overly hypertrophied and dominant, it inevitably leads to problems, which are interpreted as bureaucratization and bureaucracy.

Bureaucratization and bureaucracy in the scientific system of Kazakhstan have gradually led to a high level of incompetence and inefficiency in the administration, management, and organization of the scientific system and all science in general.

Competent and effective management is able to neutralize and mitigate some of the costs and negative narratives in the development of science. However, in our case, the incompetence and inefficiency of science management only add to the long list of costs, deficiencies, and defects in the scientific system.

Moreover, the bureaucratic style of management, lack of competencies, and absence of a scientific approach in science administration themselves become the source and generator of problems and costs for Kazakhstani science.

Let us consider bureaucratization and bureaucracy as the main causal problems of science from the perspective of Goldratt's Theory of Constraints (TOC), in the notation of Eliyahu Moshé Goldratt [Goldratt, 1997].

From the content and semantic analysis of the regulatory framework in the field of R&D, it follows that Kazakhstani regulations in the R&D sector have significant flaws, expressed in the ignorance or misunderstanding of scientific methodology and epistemology. These deficiencies are due mainly to the low level of competence of the people involved in shaping regulatory legal acts and scientific policies.

Here, we return to the problem of bureaucratization and bureaucracy in science and, consequently, to incompetence in the management system, which leads to the generally low efficiency of the scientific process in Kazakhstan.

In this context, bureaucratization and bureaucracy, as phenomenological structures and phenomena within causal relationships, are the true causes (root causes) for constructing the current reality tree (CRT) in the notation of the theory of constraints.

That is, the true cause, in turn, leads to the emergence of core problems (core problems) and, ultimately, undesirable effects (undesirable effects).

In our case, the core problem of Kazakhstani science is the phenomenon of bureaucratization and bureaucracy as a consequence of hypertrophied bureaucracy. It is bureaucratization in all its manifestations that is the root cause of undesirable effects in Kazakhstani science.

From the standpoint of causal relationships, Kazakhstani science is not an overly complex system. Even on an intuitive level, it is clear that bureaucratization and its consequence, incompetence, are the causes, whereas all other "problems" are consequences. This may not be so obvious, but we follow the logic of events.

Thus, the main problem of Kazakhstani science is the phenomenon of bureaucratization in R&D, leading to a low and noncompetitive level of quality and effectiveness of science.

Notably, bureaucracy is by no means synonymous with incompetence. There are many well-educated and knowledgeable people in R&D within a bureaucratic environment. Many of them even have academic degrees and titles. Although, of course, no academic title, insignia, or regalia can save from incompetence.

However, strictly speaking, these concepts should be distinguished. The problem is that at the decision-making level and at the HIPPO level, the most competent individuals are not selected. Thus, even within the bureaucratic environment, the competence of those generating correct proposals and solutions is undermined by the incompetence of decision-makers (HIPPO). As a result, we most often receive a negative outcome.

A vivid example of such a negative result is the full set and cycle of regulatory legal acts in the field of R&D.

Therefore, according to the theory of constraints paradigm, from the short-term and mid-term perspectives, attention should be focused on one weak link in the scientific chain—bureaucratization and bureaucracy in the national scientific system.

It is precisely B&B that acts as an attractor for the national scientific system. Bureaucratization is the main factor underlying virtually all actions, events, and initiatives in the field of science. Bureaucratization sets vectors and trajectories for the development of the entire national scientific system.

Thus, we arrive at the following important conclusions:

- Creating high-quality and progressive science is impossible both practically and theoretically without debureaucratization (eliminating or minimizing bureaucracy and bureaucratization).
- Competent individuals must be localized at the HIPPO level in the management, administration, and management of science, who are guided by and rely on a strict scientifically grounded methodology and methodology of R&D in their functions and activities.

### *The Methodology of Homeopathic Reforms*

What is the foundation of quality, effective, and productive science management, administration, and leadership?

The answer is simple—it lies in scientifically grounded methods and methodologies. That is, in correct and comprehensive project-budget documentation for scientific processing.

Throughout the years of Independence, the system of managing, administering, and leading Kazakhstani science has not given due attention to the methodological and epistemological foundations of science. These are interpreted through fields such as the science of science, epistemology, gnoseology, science studies, and scientometrics. These fields were not even formally present in our system.

Many mistakenly believe that if a person is a scientist or even a relatively prominent scientist (by our standards), they automatically know how to manage science or understand what should be done with science (the halo effect, where the more known a person is in one field, the more they are assumed to be competent in all other areas). In reality, this is not the case. In the art of management, what matters is not specific achievements but the methodology and methodology of science (as in all other fields). A scientist, or even a prominent scientist, judges science on the basis of their personal and professional experience. The problem is that this personal and professional experience, when applied by the leader (HIPPO) or science advisor, is often scaled and generalized to science as a whole. This is one of the major issues in managing science.

A mathematician does not know or understand the specifics of an archaeologist's work. A historian does not know or understand the specifics of a biochemist's work. An economist does not know or understand research in nuclear physics.

Therefore, if the experience and understanding of science from an individual authorized figure are scaled to the entire field of science, it leads to significant and hard-to-recover losses and costs in the development and progress of the entire scientific system.

This becomes especially catastrophic if one-sidedness is combined with possible personal interests.

For example, a relatively prominent scientist may propose raising the requirements for all scientists in the context of grant competition, aiming to minimize competition in conditions of financial scarcity. Alternatively, a scientist may suggest prioritizing WoS over SCOPUS when evaluating scientific projects, possibly because they publish primarily in journals indexed by WoS. The HIPPO may propose lowering publication requirements to obtain a doctoral degree with minimal effort, and so on. There are many such examples.

Naturally, all these innovations and proposals are motivated and justified. They may be made with the best intentions, for example, when the grant system is socialized (a requirement of young scientists) or when a research grant is converted into a doctoral grant.

The HIPPO sincerely tries to help young scientists by reducing the scientific *KPI* of the grant or attempting to assist doctoral students by lowering the effectiveness of the scientific grant.

Such burdensome actions make science unmanageable, and research grants lose their original purpose—of discovering and interpreting truth, novelty, or usefulness.

It is undeniable that scaling one's experience onto the entire scientific community and personal interests negatively and destructively impacts the state of science, including its prospects.

Thus, science management should not be based on voluntarism, conformism, the halo effect, the Matthew effect, or the personal ego, but rather on strict scientific methods and scientifically grounded methodologies. These should be interpreted through epistemology, gnoseology, the science of science, the philosophy of science, science studies, and scientometrics.

Therefore, this is crucial! Without methodology and the methodology of science, it is impossible to create high-quality and progressive science. Science management should be based on a solid foundation, not pulled out of thin air.

In Kazakhstan's scientific system, two structures have been created that should theoretically ensure the methodological nature of science management and the methodological justification of scientific law. These are the Department of Scientific Analytics and International Cooperation of the Ministry of Education and Science of the Republic of Kazakhstan (MNE RK) and the Analytical Center for Scientific and Scientific-Technical Development of the Academy of Sciences under the President of the Republic of Kazakhstan.

However, as it seems that they play no significant role, they are not heard of, and judging by the current legal framework for science, their goals and tasks are not being actualized. We attribute this to a lack of competence in epistemology and gnoseology, as well as a misunderstanding or underestimation of the role of epistemology and gnoseology in scientific processing.

Accordingly, we propose creating a competent institute (structure) for the methodology and methodology of science within the Institute of Philosophy, Political Science, and Religious Studies of the Science Committee of the Ministry of Science and Higher Education (MNE) (since epistemology and gnoseology are branches of philosophy) or under the auspices of the Administration of the President or the Anti-Corruption Committee. This structure should not be directly linked with the Ministry of Science and Higher Education of Kazakhstan (via a link) or the Academy of Sciences. In other words, the Administration of the President and the Anti-Corruption Committee, while observing the state and prospects of science, could be armed with a precise, objective, and professional tool in the form of such a structure institution. This would serve as a kind of scientific consulting on issues of scientific law and science management. If this structure meets expectations, its conclusions could be elevated to the HIPPO level.

Thus, scientific methodology and methodology are among the most important attributes, determinants, and tools of science policy. They serve as a kind of guideline and data-driven decision in the modern system of managing, administering, and leading science.

*Law*

Recently, almost all or the overwhelming majority of decisions within the science management system, interpreted through orders and regulatory legal acts in the field of R&D, have been met with negative reactions from the scientific community and beyond.

The most intense negative response was triggered by the latest version of the competition documentation for the years 2025--2027. The outrage was widespread, reaching open appeals (including young scientists' appeals, special opinions from individual scholars, etc.).

The majority of scientists believe that all incompetent innovations in regulatory legal acts impede the development of science, slow down scientific research, and make investments in science ineffective and unproductive.

The Anti-Corruption Committee rightly investigates the grant system for inefficient spending, conducts testing for corruption risks, and potential corruption components. However, the Anti-Corruption Committee operates on the assumption that grant funds are inefficiently and unproductively spent. This is simply a reaction to signals indicating the presence of a problem. Other supervisory and regulatory bodies often, and rightly, point to the inefficiency of grant projects.

Why is the grant system and science generally ineffective? Recall that we identified two dominant factors that are the cause of inefficiency and lack of productivity: bureaucratization of the scientific system and the relative incompetence of the science management, administration, and leadership structures.

However, the problem is that the regulatory legal framework itself largely results in inefficiency of the grant program and science in general. For example, the Ministry of Education and Science (MNE) requires efficiency and punishment for "misuse of funds." This demand is natural, correct, and justified. However, at the same time, the MNE itself creates competition documentation (CD) and other regulations that directly or indirectly hinder efficiency and force spending in nontargeted directions. Moreover, the architecture of regulatory legal acts and competition documentation clearly contributes to the extractive nature of science, particularly in terms of the survival and socialization of the scientific community.

Science is as much labor as that of a civil servant, doctor, or teacher. Any permanent work should be paid unconditionally and permanently. The ideal option would be a basic inclusive payment for scientific labor. However, given the financial deficit, this question of increasing spending on science is something we are not yet addressing.

The existing grant system is quite permanent - competitions are held annually. However, the grant system has a serious drawback: it is not inclusive and has an extractive nature. Some scientists are guaranteed grants, even several at the same time. Others do not receive grants or only receive them sporadically, interrupting or halting their scientific work. In other words, the grant system, which became dominant after the Science Law of 2011, has divided the Kazakhstani scientific community into two social groups: the salaried class and the precariat [Standing, G., 2011].

The salaried class is a social layer of researchers who have a guaranteed salary (from permanent grants), bonuses, and other privileges. The precariat is a social layer of researchers who do not have a guaranteed income (from individual discrete grants or none or receive it sporadically or randomly.

The Kazakhstani scientific salaried class includes primarily the administrative and managerial staff of scientific organizations, scientific bureaucracy, and so-called leading scientists, including those for whom the system of financial incentives and insignia is tailored. That is, in this sense, members of the Kazakhstani scientific salaried class are traditionally and officially "rewarded by trust and compensation for service" (rewarded by trust and compensation for service). We should add that this trust comes from the systemic bureaucracy and the state science governance system.

The precariat includes scientific workers with unstable scientific employment who are not part of the grant financing system or who receive grants sporadically, with a mismatch between the time spent on science and the funding received. In other words, members of the precariat have "minimal trust relationships with capital (in the state–private partnership system within the conditions of mandatory grant financing) or with the state (within R&D spending)" (minimal trust relationships with capital or the state).

At the same time, the salaried system in Kazakhstan is officially formalized—through the so-called "accreditation of subjects of scientific and scientific–technical activities," "leading scientists," commission members, etc.

However, it is unnecessary to remember that the grant system is designed specifically for scientific research, so the financial resources should be directed exclusively towards science as a whole. Grants should be inclusive for the entire scientific community.

Therefore, regulatory legal acts in the field of the grant system and science should be created not for achieving reporting goals or minimizing responsibility but for the development and progress of science itself.

Any burden unrelated to scientific processing is destructive and significantly reduces the effectiveness and productivity of scientific research.

In particular, the competition documentation includes a significant number of burdens that invalidate and serve as resistance factors for scientific progress. This is especially true for the latest competition documentation, which has effectively turned scientific grants into grants for doctoral programs or grants for the social support of young scientists. Of course, such redistribution of resources, not benefiting science itself, significantly reduces the effectiveness of science.

Therefore, the competition documentation should be subjected to revision and optimization. Any social and other obligations should be completely excluded from the competition documentation.

We do not yet specify the exact changes that need to be made to the regulatory legal framework for science. Since at this moment, we are dealing with the conceptual side.

### *Discussion*

If we successfully solve the problem of debureaucratization as the narrative of quality development, the next stage will be quantitative growth. This growth reinforces and consolidates the final stage of the transformation and modernization of the national scientific system.

Without quality development, quantitative growth has no meaning (and vice versa), as it does not lead to significant or tangible increases in the effectiveness and efficiency of science.

However, on the basis of and foundation of quality development, quantitative growth will have the expected synergistic effect—direct and rapid progress toward "high-quality and progressive science" and the country's "research hub."

We interpret quantitative growth through two main components:

- The increase in science spending to 1.2%, preferably up to 2.5%, of GDP.
- The number of scientific workers (including engineering and technical staff (ETRs) directly involved in R&D) has increased to 41,000, preferably up to 80,000.

All other forms of quantitative growth (strengthening the material and technical base, increasing the number of grant programs, expanding the range of scientific topics, expanding the number and quality of scientific publications, expanding scientific infrastructure, socialization, diversification of scientific institutions, scaling science intensity, increasing patents, enhancing innovation generation, etc.) are merely derivatives (consequences) of the two main components (cause) of quantitative growth. In other words, the parameters of other types of growth are synchronized with the dynamics and indicators of the main binary growth and are detailed separately.

The increase in science spending to 1.2% and, preferably, further to 3% of GDP.

R&D is the most crucial and effective driver of economic development.

Funding for science is funding for jobs with high labor productivity, advanced technologies, new markets, medicine, transportation, energy, ecology, and other fields.

When high labor productivity, added value, and a high standard of living are needed, the presence of science is essential. Therefore, developed countries expand the role of science in the state and increase its funding as much as possible while balancing all economic sectors. For example, since 2000, global spending on R&D has increased from $675 billion to $2.4 trillion in 2020 (Sargent, J., 2022)—a more than threefold increase in just 20 years. In the "Strategy for Industrial-Innovative Development of the Republic of Kazakhstan for 2003--2015," the following statement was reached:

*"In Kazakhstan, over the past five years, science funding has been approximately 0.2% of GDP, which is insufficient. On the basis of Kazakhstan's strategic interests, it is necessary to gradually transition to funding science at 2% of GDP by 2010 and up to 2.5%-3% by 2015."*

However, today, in 2024, 20 years later, we are still at the same level: 0.2% of GDP.

We do not need to prove again that science funding needs to be significantly increased. However, the fact remains that, for science to become a real and effective productive and technological force in the state, spending on science must reach at least 1.2% or 2.5% of GDP (the optimal level for leadership development).

National spending on science in Kazakhstan could be formed on the basis of the following financial components:

- Government subsidies (1% of GDP) (systemic science, national science, fundamental science, applied science, scientific infrastructure, the grant system);
- Sectoral internal expenses (1% of the ministry's budget) (methodology, science intensity, administrative science, science management);
- Regional scientific rent (1% of GRP) (regional science, industry-specific science, grant system);
- Corporate scientific rent (monopolies) (1% of GDP, total annual revenue) (applied science, corporate science, endowment funds, R&D commercialization);
- Resource scientific rent (subsoil users) (1% of GDP) (grant system, geonomy, geology, ecology, endowment fund);
- Stimulated private subsidies (+%) (open science, civil science, culture as science, R&D commercialization).

The main branches of science spending should be the following:

- Salaries and social security,
- Infrastructure, material-technical base (MTB), grants,
- Management (leadership), methodology, communications.

If we take the "golden ratio" of science spending, they can be interpreted as follows:

- Salaries and social security – 119 billion tenges or more (average salary of $1000 for 22,000 categories of workers)
- Infrastructure, MTB, grants – 73 billion tenges and more
- Management (leadership), methodology, communications – 45 billion tenges and more

If we increase the spending to 1.2% or better to 2.5% of GDP, with the optimal structure of spending and proportional increases in each expenditure item, we will achieve science as a full-fledged economic sector.

Once the benchmark of science spending at 1.2–2.5% of GDP is reached, it does not mean that fundamental, applied, and innovative science will immediately flood the country with high-quality results. The process of reaching authentic quality and productivity will be long and, in some cases, painful. It is associated with growing pains and overcoming complexes.

However, five to ten years from now, we will likely achieve what the President of Kazakhstan refers to as "high-quality and progressive science" and a "research hub". "

However, it is important to keep in mind that increasing spending on science is not a panacea. If the paradigms of science organization are not changed on the basis of debureaucratization, it will be yet another futile and ineffective expenditure of resources. Instead, scientific simulation and the generation of scientific and innovative simulacra will continue.

The number of scientific workers (including ETR, directly engaged in R&D) has increased to 41,000, preferably up to 80,000.
Scientific schools and groups cannot engage in scientific self-reproduction in terms of generating ideas or scientific legacy. As a result, the rapid aging of science has occurred, with no signs of growth in the quality and effectiveness of scientific research over time.

The "*Ollie effect*" (a decrease in the adaptability of a population due to its low density or number) in Kazakhstani science is further exacerbated by the requirements of competition documentation. Owing to the limited density of the scientific community, these requirements are met not through selective selection on the basis of scientific potential or talent but through random selection on the basis of social or career factors (young scientists, doctoral students). Project leaders, to secure grants, literally have to hunt for young people or doctoral students in science to meet regulatory requirements, despite these "victims" having no connection to the project's topic and often playing the role of scientific plankton or ballast.

Naturally, the Ollie effect, caused by the low density of the scientific community, leads to low productivity and quality in scientific research.

As we have mentioned, HIPPOs who approve and enact such regulations act with the best of intentions. They genuinely want to help young scientists or doctoral students in science, but in doing so, they end up harming everyone and, most of all, science itself.

The erosion and low density of scientific personnel directly hinder the reproduction of scientific potential in terms of scientific mentoring and continuity.

This issue was acknowledged by the Minister of Science and Higher Education himself:

*"We are catastrophically short of scientific personnel. At the beginning of independence, there were approximately 52,000 scientific workers, but we lost 35,000. Currently, 22,000 scientific workers are employed in research institutes and universities. Our task is now to attract youth to applied and fundamental science."*

Thus, the current number of scientific workers in the Republic of Kazakhstan is clearly insufficient for science to become an economic category, a cultural phenomenon, and a productive force.

Therefore, there are only three dominant factors in science:

- Management,
- Finance,
- Human resources.

Everything else is secondary.

Management must be competent, finances sufficient, and human resources professional. Today, this is not the case, but it must be tomorrow.

This is the foundation of our paradigm of the symbiosis of quality development and quantitative growth.

Thus, the transformation and reform of Kazakhstan's scientific system consists of two consecutive stages:

- Quality development (2025--2026)
- Quantitative growth (from 2026 onwards)

Therefore, we can state the following important fact, which we have already mentioned:

- Among all spheres of human activity, science is the most invariant and adaptive entity in relation to the structure of state governance and the political system.

As previously mentioned, science can exist and develop in any condition, environment, or ecosystem. Therefore, how science develops and progresses depends entirely on political will and state thinking.

Of course, there are social, economic, and societal limitations that set boundaries and limits for the growth and development of science. However, here, we are talking about maximizing the effect within the existing or established circumstances.

How will this manifest?

If we effectively and efficiently harness such definitions as "state thinking" and "political will," then Kazakhstan's science can become a leader in science within the post-Soviet space as a whole and in specific categories.

That is, it can become a leader in the aggregation of all scientific achievements, indicators, and metrics.

For example, in terms of quantitative indicators ("number of publications"), we objectively cannot outpace, for instance, Russian science. However, we can surpass it in qualitative indicators (citations per publication, or even in the H-index in specific fields).

Therefore, we must draw another important conclusion:

Science should and, importantly, can become a reputational symbol, brand, and foundation of the state of Kazakhstan.

Science is an important component of the state's "soft power." Through this component, Kazakhstan, limited by "hard power," can enter the leading global community.

Kazakhstan is a progressive and high-quality science, as a cultural pattern, that will generate innovations in all areas of society. In other words, to obtain innovations, we need to develop science, not innovations!

In Kazakhstan, innovations do not materialize for one simple reason: we focus on innovations while neglecting science. However, innovations stem from science.

Therefore, we must reverse the order: innovation should not lead science, science should lead innovation.

What are fundamental sciences? They are competencies.

What is applied science? It is the conversion of competencies.

What are innovations? They are the materialization of competencies.

To create a powerful innovation system, we must implement and pass through this chain:

*competencies → conversion of competencies → materialization of competencies.*

To achieve poor innovations, you need good science. To achieve good innovations, you need excellent science. In addition, to obtain excellent innovations, you need leading (cutting-edge) science.

Therefore, the existing scientific system can remain almost unchanged. However, if properly adjusted, it can not only function but also work. Implementing the listed fine tools through the successive symbiosis of quality development and quantitative growth will undoubtedly ensure positive dynamics in Kazakhstani science across all components and criteria, leading to the exit from the "*middle science trap.*"

If this concept is accepted, we will detail, chronologically, what needs to be done, which modernization must be carried out in the regulatory legal framework, what reconstructions are necessary in the management of science, and which transformations must be implemented in the functioning of the national scientific system as a whole.

That is, improve the system without changing it.

**Conclusion**

In the case of Kazakhstan, despite improvements in publication activity, the scientific system has not improved its capacity for quality discoveries. This is due to the lack of effective coordination between scientific institutions and the government, as well as insufficient integration into the international scientific community. The scientific system of Kazakhstan has faced degradation in scientific research, despite steady growth in the number of scientific workers and publications. The influence of bureaucracy on these processes cannot be ignored, as bureaucratization substitutes qualitative development of science with quantitative growth without ensuring the necessary innovation activity.

Similar problems can be observed in other Central Asian countries, where limited access to international research, internal administrative barriers, and low levels of competence in science management become the primary obstacles to the development of science in the region.

The problem of bureaucratization of science and its impact on scientific productivity remains relevant in a number of countries, including Kazakhstan, where scientific systems face the same challenges as other post-Soviet countries do. The inefficiency of state science management and excessive bureaucracy lead to a state of the "middle science trap," where the country cannot progress qualitatively in the scientific sphere despite quantitative achievements.

The key problem associated with the "middle science trap" is that the scientific system reaches a certain level of quantitative indicators (number of publications, funding) and stabilizes at that level. This is observed in Kazakhstan, where the number of scientific works is increasing, but their quality remains low, and a significant portion of the research lacks international citations. This effect is often described as "excessive quantitative growth" without qualitative progress.

Kazakhstan's scientific system is characterized not only by low-quality research but also by inefficiency in grant distribution. Government programs often face administrative obstacles that slow down the implementation of scientific projects and limit scientists' access to important international resources.

For Kazakhstan, an important step in overcoming the "middle science trap" will be the optimization of scientific processes, the improvement of the grant system, and the creation of new institutions that promote more effective use of scientific resources and the integration of science with innovative and industrial structures. To exit the "middle science trap," Kazakhstan needs to implement reforms aimed at minimizing bureaucratic barriers and creating a more flexible, open scientific system. The application of the concept of homeopathic modernization, which is integrated with international experience and considers sociocultural factors, can significantly increase the efficiency of scientific research and stimulate innovation. This will ensure sustainable growth in science and innovation, creating a foundation to overcome institutional and administrative constraints.

In general, for Kazakhstan, as well as for other transition economies, addressing the "middle science trap" and bureaucratization of the scientific system requires the integration of elements of "homeopathic modernization." The introduction of gradual changes, such as optimizing scientific processes, improving the grant system, and strengthening scientific expertise, can substantially increase the efficiency of scientific activity and help Kazakhstan reach a new level of quality. Global experience shows that such steps can significantly improve the state of scientific systems and prevent stagnation, creating opportunities for the development of innovations and increasing international competitiveness.

The concept of "homeopathic modernization" offers a way to improve the situation by focusing on minimal changes that can significantly increase the efficiency of the scientific system. The experience of countries such as China and India shows that through the strategic implementation of minimal changes, substantial progress in science and technology can be achieved.

The "homeopathic modernization" approach involves introducing minimal (point) but strategically important changes to the scientific system's structure that can eliminate key problems without causing major political and social upheavals. Unlike radical reforms, which can provoke resistance and even lead to destabilization, "homeopathic modernization" focuses on minimizing bureaucratic barriers and optimizing scientific processes, such as grant distribution, scientific project expertise, and interaction between scientists and government structures.

For Kazakhstan, the "homeopathic modernization" approach could be the optimal strategy, considering the strong bureaucratic load and low level of international competition. Issues such as excessive bureaucratization of the scientific system and an imperfect grant distribution structure remain critical barriers to scientific development. Importantly, any reforms are gradual and aimed at minimizing bureaucratic costs to increase scientific mobility and interaction between scientists, the state, and the private sector.

Kazakhstan is in the "middle science trap," where quantitative growth is not accompanied by qualitative changes. The application of the "homeopathic modernization" approach, which focuses on pinpointing changes in science management and the optimization of grant systems, could help overcome bureaucratic barriers and elevate the country to a new level of scientific competitiveness. International experience shows that such changes can significantly improve the quality of scientific research and stimulate innovation activity, which is critical for Kazakhstan's sustainable development in the global scientific and technological arena.

The "homeopathic modernization" approach suggests that, rather than radical reforms, gradual and point-specific changes should be implemented, which can lead to noticeable results in the long term. This method is especially effective in countries with strong bureaucratic systems, where large reforms could provoke resistance and lead to destabilization.

Implementing "homeopathic modernization" in Kazakhstan requires considering the specifics of the national scientific system. Key issues include low competence in science management and the lack of a scientifically grounded approach to developing regulatory legal frameworks. In particular, the current grant system does not provide equal access to resources for different categories of scientists, which contributes to growing social inequality within the scientific community.

The prospects for improving Kazakhstan's scientific system are tied to increasing funding for science, raising personnel qualifications, and implementing management methods on the basis of scientific analytics. These changes should be accompanied by minimizing bureaucratic procedures and improving interaction between scientists and management structures.

The modernization of the scientific system through optimizing grant mechanisms, reducing bureaucratic procedures, and improving scientific project expertise can significantly increase the efficiency of scientific activity without causing radical changes in the structure itself.